\documentstyle[12pt]{article}

\newcommand{\be}{\begin{equation}}
\newcommand{\ee}{\end{equation}}
\newcommand{\dlt}{\delta}
\newcommand{\Dlt}{\Delta}
\newcommand{\ra}{\rightarrow}
\newcommand{\vp}{\varphi}
\newcommand{\bt}{\beta}
\newcommand{\al}{\alpha}
\newcommand{\prt}{\partial}

\newcommand{\om}{\omega}
\newcommand{\Lbd}{\Lambda}

\newcommand{\ep}{\varepsilon}

\newcommand{\br}{{\bf r}}
\newcommand{\bB}{{\bf B}}

\newcommand{\bj}{{\bf j}}
\newcommand{\bA}{{\bf A}}
\newcommand{\ba}{{\bf a}}
\newcommand{\dgr}{\dagger}

\begin{document}

\begin{center}
{\Large{\bf Multiple Coupling of Topological Coherent Modes
of Trapped Atoms} \\ [5mm]

V.I. Yukalov$^{1,2}$, K.-P. Marzlin$^{2,3}$, and E.P. Yukalova$^4$} \\ [5mm]

{\it
$^1$Bogolubov Laboratory of Theoretical Physics \\
Joint Institute for Nuclear Research, Dubna 141980, Russia\\ [3mm]

$^2$Fachbereich Physik, Universit\"at Konstanz \\
Fach M674, 78457 Konstanz, Germany \\ [3mm]

$^3$Department of Physics and Astronomy\\
2500 University Drive NW\\
Calgary, Alberta T2N 1N4, Canada\\[3mm]

$^4$Department of Computational Physics \\
Laboratory of Information Technologies \\
Joint Institute for Nuclear Research, Dubna 141980, Russia}

\end{center}

\vskip 2cm

\begin{abstract}

The possibility of generating multiple coherent modes in trapped Bose
gases is advanced. This requires the usage of several driving fields
whose frequencies are tuned close to the corresponding transition
frequencies. A general criterion is derived explaining when the
driving fields, even being in perfect resonance, cannot generate the
topological coherent modes. This criterion is termed the theorem of
shape conservation. Bose-Einstein condensates with generated coherent
modes display a number of interesting effects, such as: interference
fringes, interference current, mode locking, dynamic transition,
critical phenomena, chaotic motion, harmonic generation, parametric
conversion, atomic squeezing, and entanglement production.
Approximate solutions, based on the averaging techniques, are
found to be in good agreement with direct numerical calculations
for the Gross-Pitaevskii equation.

\end{abstract}

\newpage

\section{Introduction}

Dilute Bose gases at low temperatures are described by the
Gross-Pitaevskii equation (see reviews [1--6]). The ground-state
solution of the stationary Gross-Pitaevskii equation corresponds
to the equilibrium Bose-Einstein condensate. In the presence of
a trapping potential, this equation possesses a discrete spectrum
and, respectively, a discrete set of wave functions describing
the topological coherent modes, which represent nonground-state
condensates [7]. These modes are called topological since the
related wave functions have different spatial shapes, with
different number of zeros. And the modes are termed coherent
because the related wave functions correspond to coherent states [8].
Vortices are a particular case of such modes [9,10].

In practice, the topological coherent modes can be generated by
subjecting the system of trapped atoms to the action of an
alternating external field with a frequency tuned to the resonance
with a transition frequency between the ground state and an excited
collective level [7,9--12]. The properties of these modes have been
studied in several publications [7--26]. Instead of modulating with
a resonant field the trapping potential, one can periodically vary
the atomic scattering length by invoking the Feshbach resonance
[27--29]. Note that the coherent modes, being the solutions to the
nonlinear Gross-Pitaevskii equation, in some cases can be considered
as analytical continuations, under increasing nonlinearity, of the
related linear modes; but there exist also the irreducible modes
having no linear counterparts. The latter happens for sufficiently
strong nonlinearity due to atomic interactions, when atoms are in
double- or multiwell potentials [19,22] or inside an optical lattice
[30--33]. Thermodynamics of the coherent modes has also been considered [34].

One should not confuse the topological coherent modes, which are
the solutions to the nonlinear Gross-Pitaevskii equation, with the
elementary excitations treated by the linear Bogolubov - De Gennes
equations. In principle, one could consider resonant transitions
between the modes of elementary excitations [35], which, however,
is a radically different problem.

In previous publications, the case of a single generated coherent mode
has been studied, which requires the action of one external resonant
field. Here, we generalize the consideration to the multiple mode
generation, which can be done by involving several alternating
fields. It is particularly important that we have accomplished
all investigations in two ways, by employing approximate analytical
methods, based on the averaging techniques, and also by solving
numerically the Gross-Pitaevskii equation, both ways being in good
agreement with each other.

\section{Multiple Mode Generation}

Topological coherent modes are the solutions to the stationary
Gross-Pitaevskii equation
\be
\label{1}
\hat H [\vp_n]\vp_n(\br) = E_n\vp_n(\br) \; ,
\ee
with the nonlinear Hamiltonian
\be
\label{2}
\hat H[\vp] = -\; \frac{\hbar^2}{2m_0}\; \nabla^2 + U(\br) +
NA_s|\vp|^2
\ee
containing a trapping potential $U(\br)$ and the nonlinear term due
to the binary contact interactions
$$
A_s \equiv 4\pi\hbar^2\; \frac{a_s}{m_0} \; ;
$$
$N$ being the number of particles in the trap; $m_0$, particle mass;
$a_s$, scattering length. The functions $\vp_n(\br)$ are normalized
to unity, $(\vp_n,\vp_n)=1$. In general, $\vp_m$ and $\vp_n$ are not
compulsorily orthogonal.

Temporal evolution is described by the time-dependent Gross-Pitaevskii
equation
\be
\label{3}
i\hbar \; \frac{\prt}{\prt t} \; \vp(\br,t) = \left ( \hat H[\vp] +
\hat V\right ) \vp(\br,t) \; ,
\ee
with an external potential $\hat V=
V(\br,t)$, which is assumed to have the multifrequency form
\be
\label{4}
\hat V = \frac{1}{2}\; \sum_j \left [ B_j(\br) e^{i\om_j t} + B_j^*(\br)
e^{-i\om_j t} \right ] \; ,
\ee
in which $j=1,2,\ldots$. The frequencies of the alternating potential (4)
are chosen so that each of them is tuned close to one of the transition
frequencies
\be
\label{5}
\om_{mn} \equiv \frac{1}{\hbar} \left ( E_m - E_n \right )
\ee
between two coherent modes, which implies the validity of the resonance
conditions
\be
\label{6}
\left | \frac{\Dlt_{mn}}{\om_{mn}}\right | \ll 1 \; , \qquad
\Dlt_{mn} \equiv \om_j - \om_{mn} \; .
\ee

Transitions between different modes can be produced by the action
of the multiresonant potential (4) and also owing to interatomic
interactions. Consequently, there are two types of transition
amplitudes, defined as the matrix elements over the related coherent
modes. The transition amplitudes
\be
\label{7}
\al_{mn} \equiv N\; \frac{A_s}{\hbar} \left ( |\vp_n|^2,\;
2|\vp_n|^2 - |\vp_m|^2 \right )
\ee
correspond to interatomic interactions, while the transition
amplitudes
\be
\label{8}
\bt_{mn} \equiv \frac{1}{\hbar} \left ( \vp_m,\hat B_j\vp_n
\right ) \; ,
\ee
with $\hat B_j=B_j(\br)$, are characterized by the related resonant
parts of the potential (4). Good resonance can be supported
provided the transition amplitudes (7) and (8) are much smaller than
the related transition frequencies,
\be
\label{9}
\left | \frac{\al_{mn}}{\om_{mn}} \right | \ll 1 \; , \qquad
\left | \frac{\bt_{mn}}{\om_{mn}} \right | \ll 1 \; .
\ee
The first of these inequalities limits the number of atoms in the
trap [12]. This limiting number of atoms is of the order of the
critical number of atoms with attractive interactions allowing
for the stability of the trapped atomic gas [7,12,36]. Hence,
the generation of coherent modes is admissible for atomic systems
with repulsive as well as attractive interactions.

The solution to Eq. (3) can be presented as the expansion
\be
\label{10}
\vp(\br,t) = \sum_n c_n(t) \vp_n(\br) \;
\exp\left ( -\; \frac{i}{\hbar}\; E_n t\right )
\ee
over the coherent modes. The coefficient functions $c_n(t)$ are
assumed to be slowly varying in time as compared to the exponentials,
so that
\be
\label{11}
\frac{1}{\om_{mn}} \; \left | \frac{dc_n}{dt} \right | \ll 1 \; .
\ee
Inequalities (8) and (11) are mutually self-consistent, since if
$\al_{mn}\ra 0$ and $\bt_{mn}\ra 0$, then $c_n(t)\ra const$ and
$dc_n/dt\ra 0$. Substituting expansion (10) into the evolution
equation (3) and employing the averaging technique [7,12] results
in the system of equations for the functions $c_n(t)$. The values
$|c_n(t)|^2$ describe fractional mode populations. For instance,
in the case of three modes coupled by two resonant fields, we find
$$
i\; \frac{dc_1}{dt} = \left ( \al_{12} |c_2|^2 + \al_{13}|c_3|^2
\right ) c_1 + f_1 \; ,
$$
$$
i\; \frac{dc_2}{dt} = \left ( \al_{21} |c_1|^2 + \al_{23}|c_3|^2
\right ) c_2 + f_2 \; ,
$$
\be
\label{12}
i\; \frac{dc_3}{dt} = \left ( \al_{31} |c_1|^2 + \al_{32}|c_2|^2
\right ) c_3 + f_3 \; ,
\ee
where the terms $f_j$ depend on the kind of the mode generation
involved. Thus, for the cascade generation,
$$
f_1=\frac{1}{2}\; \bt_{12} c_2 e^{i\Dlt_{21}t} \; , \qquad
f_2=\frac{1}{2}\; \bt_{12}^* c_1 e^{-i\Dlt_{21}t}
+ \frac{1}{2}\; \bt_{23} c_3 e^{i\Dlt_{32}t} \; ,
$$
\be
\label{13}
f_3=\frac{1}{2}\; \bt_{23}^* c_2 e^{-i\Dlt_{32}t} \; ;
\ee
for the $V$-type generation,
$$
f_1=\frac{1}{2}\; \bt_{12} c_2 e^{i\Dlt_{21}t}
+ \frac{1}{2}\; \bt_{13} c_3 e^{i\Dlt_{31}t} \; ,  \qquad
f_2=\frac{1}{2}\; \bt_{12}^* c_1 e^{-i\Dlt_{21}t} \; ,
$$
\be
\label{14}
f_3=\frac{1}{2}\; \bt_{13}^* c_1 e^{-i\Dlt_{31}t} \; ;
\ee
and for the $\Lbd$-type generation
$$
f_1=\frac{1}{2}\; \bt_{13} c_3 e^{i\Dlt_{31}t} \; , \qquad
f_2=\frac{1}{2}\; \bt_{23} c_3 e^{i\Dlt_{32}t}  \; ,
$$
\be
\label{15}
f_3=\frac{1}{2}\; \bt_{13}^* c_1 e^{-i\Dlt_{31}t} +
\frac{1}{2}\; \bt_{23}^* c_2 e^{-i\Dlt_{32}t}\; .
\ee
Equations (12) are to be complimented by initial conditions and by
the normalization condition
$$
|c_1|^2+|c_2|^2+|c_3|^2  = 1 \; .
$$
Note that Eqs. (12) enjoy the global phase symmetry $c_j\ra c_j\exp(i\chi)$,
with $\chi\in[0,2\pi]$. This could allow us to take one of the initial
conditions for $c_j(0)$ as a real quantity. However, one cannot make
all three functions $c_j(t)$ real. In general, they are complex-valued
functions.

Let us emphasize that all generated modes coexist in the same trap
and are not spatially separated. This distinguishes the considered
situation, both physically as well as mathematically, from the cases
of spatially separated condensates, as in multiwell potentials [37]
or in clouds separated by means of the collective Rayleigh scattering
[38--41]. The coherent modes, we consider, differ from each other by
the shapes of their wave functions and by their energies, which are
defined by the Gross-Pitaevskii equation (1).

\section{Theorem of Shape Conservation}

Applied alternating potentials not always can generate new topological
coherent modes, even if the required resonance conditions are perfectly
valid. It may occur that the initial atomic cloud, being subject to the
action of resonant fields, oscillates in space without changing its
shape, but no higher modes are excited. The criterion for such
a behaviour is provided by the theorem below.

Let us consider Bose-condensed atoms in a trapping potential $U(\br)$,
with a driving field $V(\br,t)$. The corresponding coherent wave
function $\vp(\br,t)$ satisfies the temporal Gross-Pitaevskii equation
(3). Since atoms are trapped, then the {\it trapping condition}
\be
\label{16}
\vp(\br,t)\ra 0 \qquad (|\br|\ra\infty)
\ee
is always valid for all $t\geq 0$. Assume that at the initial time the
function $\vp(\br,0)$ represents a real mode
\be
\label{17}
\vp(\br,0) =\vp_0(\br) =\vp_0^*(\br)
\ee
satisfying the stationary equation (1). We shall say that the density
of atoms conserves its shape when the {\it shape-conservation condition}
\be
\label{18}
|\vp(\br,t)| = |\vp(\br-\ba,0)|
\ee
holds true, where $\ba=\ba(t)$ defines the center-of-mass motion. The
following criterion has been proven [42].

\vskip 2mm

{\bf Theorem}. The solution to the Gross-Pitaevskii equation (3), obeying
conditions (16) and (17), satisfies the shape-conservation condition (18)
if and only if the trapping potential is harmonic, while the driving
field is linear in space, that is,
\be
\label{19}
U(\br) =A_0 +\bA_1\cdot\br +\sum_{\al\bt} A_{\al\bt} r^\al r^\bt \; ,
\qquad V(\br) =B_0(t) +\bB_1(t)\cdot\br \; ,
\ee
where $B_0(t)$ and $\bB_1(t)$ are arbitrary functions of time. Then the
center-of-mass motion is given by the equation
\be
\label{20}
m_0\; \frac{d^2a^\al}{dt^2}  + \sum_\bt (A_{\al\bt} + A_{\bt\al} )
a^\bt +B_1^\al(t) = 0 \; .
\ee

This theorem shows that, even when the functions $B_0(t)$ and
$\bB_1(t)$ in Eq. (19) correspond to alternating fields being in
resonance with some transition frequencies, no generation of other
modes is possible, but the density of atoms will conserve its shape,
with the center-of-mass motion described by Eq. (20). Hence, in order
to effectively generate coherent modes, one has to avoid condition (19).

\section{Dynamic Resonant Effects}

By applying $k$ alternating fields, with the frequencies $\om_j$
$(j=1,2,\ldots,k)$ one can generate $k$ topological modes, starting
with one given mode, say, having at the initial time all atoms in the
ground state. Then we will get $k+1$ coupled coherent modes. The mode
amplitudes $c_j$, defining the fractional populations $|c_j|^2$, obey
$k+1$ nonlinear differential equations. Because of their nonlinearity,
these equations display a rich variety of interesting features. We
have thoroughly examined the dynamics of $c_j$ for the cases of two
and three coupled coherent modes, finding all fixed points and
accomplishing a complete stability analysis. For the corresponding
physical cases, we have also solved the Gross-Pitaevskii equation
numerically. Both approaches were found to be in good agreement with
each other. Since the derivation of the equations for $c_j$ is based
on the averaging technique [43], the resulting equations are approximate,
with the accuracy of their solutions characterized by the values of
the involved parameters. The errors of such approximate solutions are
of the order of $\max\{|\al_{mn}/\om_{mn}|,|\bt_{mn}/\om_{mn}|\}$.
Within this accuracy, the results obtained by solving the evolution
equations for the mode amplitudes $c_j$ coincide with those following
from the direct numerical solution of the Gross-Pitaevskii equation.
Details of these calculations will be published in a separate paper.
And here, being limited in space, we shall present a brief account
of the most interesting physical effects we have found.

\vskip 3mm

(1) {\it Interference Fringes}

\vskip 2mm

The total density of atoms in the trap, $\rho(\br,t)\equiv|
\vp(\br,t)|^2$, is not equal to the sum of the partial densities
$\rho_j(\br,t)\equiv|\vp_j(\br,t)|^2$ where $\vp_j(\br,t)=c_j(t)\vp_j(\br)$,
but there exist the characteristic interference fringes described by the
interference density
\be
\label{21}
\rho_{int}(\br,t) \equiv \rho(\br,t) - \sum_j \rho_j(\br,t) \; .
\ee
The appearance of such interference patterns is typical of coexisting
coherent states [12,20,25,26].

\vskip 3mm

(2) {\it Interference Current}

\vskip 2mm

Because of an essential nonuniformity of the atomic density for
several coexisting coherent modes, there arises the interference
current
\be
\label{22}
\bj_{int}(\br,t) \equiv \bj(\br,t) - \sum_i \bj_i(\br,t) \; ,
\ee
which, similarly to the interference density (21), is the difference
between the total current
$$
\bj(\br,t) \equiv \frac{\hbar}{m_0} \; {\rm Im}\; \vp^*(\br,t)
\vec\nabla\vp(\br,t)
$$
and the sum of the partial currents
$$
\bj_i(\br,t) \equiv \frac{\hbar}{m_0}\; {\rm Im}\; \vp_i^*(\br,t)
\vec\nabla \vp_i(\br,t) \; .
$$
The interference current is also typical of the case of different
coexisting coherent modes [12,20,25,26]. Such a current is called
sometimes the internal Josephson current or topological current.
The possibility of the Josephson-type oscillations between two
interpenetrating populations, not separated by any  barrier, was
suggested by Leggett [44].

\vskip 3mm

(3) {\it Mode Locking}

\vskip 2mm

When the amplitudes of the driving resonant fields are sufficiently
small, such that $|\bt_{mn}/\al_{mn}|\ll 1$, the fractional mode
populations exhibit nonlinear oscillations close to their initial
values, never crossing the line $1/2$, that is, depending on initial
conditions, one has either
\be
\label{23}
0\leq |c_j|^2 \leq \frac{1}{2}
\ee
or
\be
\label{24}
\frac{1}{2} \leq |c_j|^2 \leq 1 \; .
\ee
The mode locking effect exists for two [7,11,12] as well as for several
[42] coupled modes. Mathematically, it is the same as self-trapping [45]
of atoms in one of the wells of a stationary multiwell potential.

\vskip 3mm

(4) {\it Dynamic Transition}

\vskip 2mm

Increasing the amplitudes of the driving fields up to the values such
that $|\bt_{mn}/\al_{mn}|\approx 0.5$, one passes from the mode-locked
regime of motion to the mode-unlocked regime, when the mode populations
oscillate in the whole region between zero and one,
\be
\label{25}
0\leq |c_j|^2 \leq 1 \; .
\ee
The change of this dynamic behaviour occurs when the starting point of
a trajectory is crossed by a saddle separatrix [11,12,25,26]. A similar
effect of the change from the self-trapped to untrapped regime exists
for multiwell potentials [46].

\vskip 3mm

(5) {\it Critical Phenomena}

\vskip 2mm

On the manifold of the system parameters $\al_{mn},$ $\bt_{mn}$, and
$\Dlt_{mn}$, there exist surfaces whose crossing results in the dynamic
transitions from the mode-locked to mode-unlocked regimes of motion.
In the vicinity of these separating surfaces, the temporal behaviour
of the mode populations experiences sharp changes [11,12,18,20,26],
because of which the separating surface can be termed the critical
surface. Moreover, the time-averaged system exhibits on a critical
surface critical phenomena analogous to those happening in equilibrium
statistical systems under second-order phase transitions [20].

\vskip 3mm

(6) {\it Chaotic Motion}

\vskip 2mm

When the amplitudes of the driving fields become so large that the
ratio $|\bt_{mn}/\al_{mn}|$ is of order one, the temporal evolution
of the fractional mode populations $|c_j|^2$ can go chaotic. This
takes place only when the number of the coupled coherent modes is
equal or larger than three. Two coexisting modes display always
a regular behaviour [11,12,26]. The onset of chaos for three or more
modes is connected with the disappearance of stable fixed points [42].
This is similar to the arising chaos in a three-well potential [37].

\vskip 3mm

(7) {\it Harmonic Generation}

\vskip 2mm

In order to generate a coherent mode, it is not compulsory to invoke
the standard resonance, as in Eq. (6), when the frequency $\om$ of the
driving field is close to the required transition frequency $\om_{21}=
(E_2-E_1)/\hbar$. But, instead of the resonance condition $\om=\om_{21}$,
one can employ one of the harmonic generation conditions
\be
\label{26}
n\om =\om_{21} \qquad (n=1,2,\ldots ) \; .
\ee
This type of harmonic generation is well known in quantum optics [47]
and can also be realized for elementary excitations in Bose-Einstein
condensates [48,49]. As we have shown [42], it exists as well for
topological coherent modes.

\vskip 3mm

(8) {\it Parametric Conversion}

\vskip 2mm

When several driving fields are involved, with the frequencies $\om_j$
($j=1,2,\ldots$), then there exist other possibilities for generating
a coherent mode corresponding to the transition frequency $\om_{21}$.
Thus, in the case of two driving fields, with the frequencies $\om_1$
and $\om_2$, the transition from a mode with energy $E_1$ to that of
energy $E_2$ can be activated if $\om_1+\om_2=\om_{21}$ or $\om_1-\om_2=
\om_{21}$. In the language of quantum optics [47], this is called
parametric conversion, up or down, respectively. Generally, for several
driving fields, the condition of parametric conversion can be written
as
\be
\label{27}
\sum_j (\pm\om_j) = \om_{21} \; .
\ee
An analogous effect is known in quantum optics [47] and for elementary
excitations in trapped condensates [48,49]. Now, we proved [42] its
existence for nonlinear coherent modes.

\vskip 3mm

(9) {\it Atomic Squeezing}

Atomic squeezing can be defined by means of pseudospin operators, because
of which it is often termed spin squeezing. For this purpose, one passes
from the atomic operators $a_i$ and $a_i^\dgr$, corresponding to the $i$-th
mode and satisfying the commutation relations $[a_i,a_j^\dgr]=\dlt_{ij}$,
to the pseudospin operators. In the case of the two-mode condensate, with
$i=1,2$, one defines the collective spin operators
$$
S_- = a_1^\dgr a_2 \; , \qquad S_+ = a^\dgr_2 a_1 \; , \qquad
S_z = \frac{1}{2}\left ( a_2^\dgr a_2 - a_1^\dgr a_1 \right ) \; ,
$$
which satisfy the standard spin commutation relations
$$
[S_+,S_-]=2S_z \; , \qquad [S_z,S_\pm] = \pm S_\pm \; .
$$
The amount of squeezing can be measured by the squeezing factor
\be
\label{28}
Q_S \equiv \frac{2\Dlt^2(S_z)}{|<S_\pm>|} \; ,
\ee
where $\Dlt^2(S_z)$ is the dispersion. Atomic operators are connected
with the mode amplitudes by the equality $<a_i^\dgr a_j>=Nc_i^*c_j$.
Since $c_i=c_i(t)$ is a function of time, the squeezing factor (28)
is an oscillating function [12].

For the 3-mode condensate, when $i=1,2,3$, one introduces the collective
angular momentum operators
$$
J_- =\sqrt{2}\left ( a_1^\dgr a_2 + a_2^\dgr a_3 \right ) \; , \qquad
J_+ =\sqrt{2}\left ( a_2^\dgr a_1 + a_3^\dgr a_2 \right ) \; , \qquad
J_z =  a_3^\dgr a_3 - a_1^\dgr a_1 \; ,
$$
with the commutation relations
$$
[J_+,J_-] = 2J_z \; , \qquad [J_z,J_\pm]=\pm J_\pm \; .
$$
Then squeezing can be characterized by the squeezing factor
$$
Q_J \equiv \frac{2\Dlt^2(J_z)}{|<J_\pm>|} \; ,
$$
similar to factor (28). The generation of atomic squeezing can be
useful for spectroscopy.

\vskip 3mm

(10) {\it Entanglement Production}

\vskip 2mm

To quantify the amount of entanglement produced by a $p$-particle
density matrix $\rho_p(t)$, we may employ the measure
\be
\label{29}
\ep(\rho_p) =\log\;
\frac{||\rho_p||_{\cal D}}{||\rho_p^\otimes||_{\cal D}} \; ,
\ee
where $||\rho_p||_{\cal D}$ implies the Hermitian norm over a set
${\cal D}$ of disentangled states and $\rho_p^\otimes$ means the
disentangled counterpart of $\rho_p$ (see details in [50]). For a
multimode condensate, we have
\be
\label{30}
\ep(\rho_p) = (1-p) \log\; \sup_j |c_j|^2 \; .
\ee
The mode amplitudes $c_j=c_j(t)$ depend on time. Therefore measure (30)
describes the evolutional entanglement. The temporal behaviour of the
functions $c_j(t)$ can be regulated by switching on and off the resonant
driving fields. Hence the measure (30) of the produced entanglement can
also be regulated, which opens the possibility for information processing
by means of the multimode resonant condensates.

\section{Conclusion}

We have shown that multiple topological coherent modes can be generated
in a Bose-Einstein condensate, which requires the usage of several
driving fields. It is important to emphasize that we have accomplished
the consideration in two ways, by applying the averaging techniques and
by direct numerical simulations for the Gross-Pitaevskii equation, both
ways being in good agreement with each other. The multimode condensate
presents a novel object possessing a variety of interesting properties.
From one side it reminds us a multilevel resonant atom, so widely
considered in quantum optics, but, at the same time, it is a multiatomic
system, because of which it exhibits unusual nonlinear features. The
multimode condensate displays the following remarkable effects:
interference fringes, interference current, mode locking, dynamic
transition, critical phenomena, chaotic motion, harmonic generation,
parametric conversion, atomic squeezing, and entanglement production.
The rich variety of interesting properties, characteristic of the
multimode condensates, suggests that the latter could find a number of
applications.

\vskip 5mm

{\bf Acknowledgement}

\vskip 2mm

We acknowledge financial support from the Heisenberg-Landau Program,
Forschergruppe Quantengase, and the Optik Zentrum Konstanz.

\end{document}